\documentclass[smallextended]{svjour3}

\usepackage{graphicx}
\usepackage{amsmath}
\usepackage{amsfonts}
\usepackage{epstopdf}
\usepackage{ulem}
\usepackage{color}

\smartqed

\begin{document} 

\title{Motility-induced phase separation of active particles in the presence of velocity alignment}
%%  of Can interacting self-propelled particle systems be mapped into equilibrium systems?} 
% \\ Derivation of an entropy functional for interacting self propelled particles in the disordered phase}

%author{J. Barr\'e$^{1}$, R. Ch\'etrite$^{1}$, M.Muratori$^{1,2}$, F. Peruani$^{1}$\\
%    ${}^{1}$ Laboratoire J.A. Dieudonn\'e, Universit\'e de Nice Sophia-Antipolis,\\
%    UMR CNRS 7351, Parc Valrose, F-06108 Nice Cedex 02, France\\
%    ${}^{2}$ INRIA...}
\author{Julien Barr\'e \and Rapha\"el Ch\'etrite \and Massimiliano Muratori \and Fernando Peruani}

\institute{J. Barr\'e, R. Ch\'etrite, M. Muratori and F. Peruani  \at Universit\'e Nice Sophia Antipolis, Laboratoire J.A. Dieudonn\'e, UMR CNRS 7351, Parc Valrose, F-06108 Nice Cedex 02, France
}
%\author{J. Barr\'e} 
%\affiliation{Universit\'e Nice Sophia Antipolis, Laboratoire J.A. Dieudonn\'e, UMR CNRS 7351, Parc Valrose, F-06108 Nice Cedex 02, France}
%\author{R. Ch\'etrite} 
%\affiliation{Universit\'e Nice Sophia Antipolis, Laboratoire J.A. Dieudonn\'e, UMR CNRS 7351, Parc Valrose, F-06108 Nice Cedex 02, France}
%\author{M. Muratori}
%\affiliation{Universit\'e Nice Sophia Antipolis, Laboratoire J.A. Dieudonn\'e, UMR CNRS 7351, Parc Valrose, F-06108 Nice Cedex 02, France}
%\author{F. Peruani}
%\affiliation{Universit\'e Nice Sophia Antipolis, Laboratoire J.A. Dieudonn\'e, UMR CNRS 7351, Parc Valrose, F-06108 Nice Cedex 02, France}

\begin{abstract}
Self-propelled particle (SPP) systems are intrinsically out of equilibrium systems, where each individual particle converts energy into work to move in a dissipative medium. 
%
% that exhibit energy consumption as well as dissipation. 
%
When interacting through a velocity alignment mechanism, and with the medium acting as a momentum sink, even momentum is not conserved. 
In this scenario, a mapping into an equilibrium system seems unlikely. 
Here, we show that an entropy functional can be derived for SPPs with velocity alignment and density-dependent speed, at least in the (orientationally) disordered phase. 
This non-trivial result has important physical consequences. 
The study of the entropy functional reveals that the system can undergo phase separation  before the orientational-order phase transition known to occur in SPP systems with velocity alignment.
% , i.e., in the disordered phase. 
%
Moreover, we indicate that the spinodal line is a function of the alignment sensitivity and show that 
density fluctuations as well as the critical spatial diffusion, that leads to phase separation, dramatically increase as  the orientational-order transition is approached. 
% 
%% , diverging at the point where the orientational-order phase transition occurs.  
%% We show that density fluctuations also diverge as the orientational-order critical point is approached.
%
%% the analysis shows that phase-separation critical point is controlled by the alignment strength, even before any global orientational order is  observed. 
%
%% These observations are inline with  recent experimental and simulations results.
%
\end{abstract}

\maketitle

\section{Introduction}

Examples of biological interacting self-propelled particle (SPP) systems include animal groups~\cite{cavagna10,bhattacharya10}, 
insect swarms~\cite{buhl06,romanczuk09}, bacteria~\cite{zhang10,peruani2012,starruss2012}, and,  
at the microcellular scale, microtubules driven by molecular motors~\cite{schaller10}.
Even though most examples of SPPs  come from biology, there exist non-living SPP systems. 
There are several ways of fabricating artificial SPPs.  The self-propulsion of such particles typically requires an asymmetry in the particle:  two distinct friction coefficients~\cite{kudrolli2008,deseigne2010,weber2013}, 
light absorption  coefficients~\cite{jiang2010,golestanian2012,theurkauff2012,palacci2013}, or catalytic properties~\cite{paxton2004,mano2005,rucker2007,howse2007,golestanian2007}  depending on whether energy injection is done through vibration, 
light emission, or chemical reaction, respectively. 
Interestingly, this asymmetry does not need to be an intrinsic particle property.
 Self-propelled Quincke rollers~\cite{bartolo2013} as well as actively moving drops~\cite{thutupalli2011} are  remarkable examples where 
the asymmetry results from a spontaneous symmetry breaking that sets the particle to move in a given direction.

At a theoretical level, we have learned in the recent years that the large-scale properties of SPP systems depend on few microscopic details. 
The symmetry associated to the self-propulsion mechanism of the particles, which can be either {\it 
polar}~\cite{vicsek1995,gregoire2004,peruani2008} or {\it apolar}~\cite{chate2006},  
as well as the symmetry of the particle-particle interactions, that often occur via a velocity alignment mechanism, 
which  can be either {\it ferromagnetic}~\cite{vicsek1995,gregoire2004}   
or {\it nematic}~\cite{chate2006,peruani2008,peruani2006}, play a fundamental role in the resulting self-organized patterns. 
Equally important is the dimension of the space where particles move, whether this space is continuous (off-lattice)~\cite{vicsek1995,gregoire2004,chate2006,peruani2008} or discrete (on lattice)~\cite{bussemaker1997,csahok1995,loan1999,raymond2006}, and whether particles move on a homogeneous or heterogeneous medium~\cite{chepizkho2013a,chepizkho2013b,reichhardt2014,quint2014}.

Another aspect of key importance, and central to the present study, is whether there exists a coupling between the speed and the density  of the SP particles. 
Notice that the importance lies on the existence of such coupling and not on the mere fact that the speed may fluctuate. 
In the context of SPPs with a velocity alignment, it has been shown first with a lattice model~\cite{peruani2011} and later on with an off-lattice model~\cite{tailleur2012} 
that such a coupling induces spontaneous phase separation and a zoology of complex patterns. 
The most evident physical mechanism that can introduce a coupling between speed and density is simple volume exclusion  as showed with  simple lattice models by  adding exclusion process rules  in the absence of particle-particle alignment in~\cite{thompson2011} and with alignment in~\cite{peruani2011}, and with an off-lattice model of self-propelled disks interacting by a soft-core repulsion~\cite{fily2012}. 
The observed non-equilibrium phase separation can be traced back to the non-equilibrium {\it Motility Induced Phase Separation} (MIPS) 
introduced in the context of interacting run-and-tumble particles by J. Tailleur and M.E. Cates in~\cite{tailleur2008,tailleur2013}. 
In absence of an alignment mechanism, MIPS is a generic feature of active particles interacting by volume exclusion as shown in simulations with 
self-propelled disks~\cite{fily2012,redner2013,fily2014} and spheres~\cite{valeriani2013,wysocki2013}, and argued theoretically in~\cite{thompson2011,cates2013,speck2013}. 
One exciting aspect of the MIPS, as first pointed out in~\cite{thompson2011,tailleur2008,tailleur2013}, is the remarkable similarity with equilibrium phase-separation, 
which allows the mapping between these non-equilibrium active systems with the analogous equilibrium systems. 

The goal of the present study is to look at MIPS in the  context of SPPs with a velocity alignment mechanism. 
Specifically, we want to understand the role played by the alignment mechanism in the phase separation process. 
Let us recall that SPP systems with a velocity alignment mechanism exhibit a phase transition from a disordered to an ordered phase. 
In the disordered phase, the large-scale behavior of the particles is diffusive as occurs for SPPs without a velocity alignment. 
Thus, we may hope that a mapping to an equilibrium scenario, as the one performed in~\cite{tailleur2008,tailleur2013}, remains possible.  
%
%For SPP systems without an velocity alignment mechanism, a mapping to equilibrium systems is built in 
%\cite{tailleur2008,tailleur2013}. With alignment, SPP systems exhibit an 
%ordered and a disordered phase; in the disordered phase, the large scale 
%behavior of the particles is diffusive, as in the case without alignment. 
%Thus, we may hope that a similar mapping to equilibrium remains possible. }
%
%We stress that the existence of such an alignment mechanism prevents us from making a direct and obvious mapping to equilibrium systems. 
%
%Despite of this, since SPP systems exhibit an ordered and a disordered phase, we can expect that the inside the disordered phase such a mapping is still possible. 
%
We push for such an analogy as far as possible. To be exact, to the onset of the ordered phase. 

Before starting, let us review briefly some of the most relevant theoretical results for (dried) SPP systems with velocity alignment and in the absence of density-dependent speed. 
The first hydrodynamical equations were derived based on symmetry arguments and contained all allowed terms by symmetry~\cite{toner1995,toner1998}.
These initial studies provided a theoretical basis to understand the emergence of long-range order (LRO) in two dimensional systems with continuum symmetry 
as well as the presence of giant number fluctuations in the ordered phase. 
The drawback of these initial approaches is the impossibility of connecting the parameters of the hydrodynamic equations with those of the microscopic models. 
In~\cite{bertin2009}, the macroscopic equations were derived from given microscopic equations. 
Such an approach revealed that the ``parameters'' of the hydrodynamic theory are in fact non-linear functions of the density. 
This has  allowed to understand the emergence of macroscopic structures, such as bands, in this type of SPP systems~\cite{mishra2010,caussin2014}. 
%,  which have  been agued to be generic features of these systems~\cite{mishra2010, caussin2014}. 
%
For a detailed review,  we refer the reader to~\cite{marchetti2013}. 
%% on the derivation of hydrodynamic equations (without fluctuating terms) for SPP systems, we refer the reader to~\cite{marchetti2013}. 
Here, we just mention that macroscopic equations have been derived for ferromagnetic~\cite{toner1995,toner1998,bertin2009,marchetti2013} and nematic velocity alignment~\cite{peshkov2012}   
in the dilute approximation and close to the order-disorder transition, with the exception of~\cite{ihle2011} and~\cite{degond2008}. 
Even though we have now a fairly good qualitative understanding of the hydrodynamics of SPP systems (in homogeneous media), many open questions and fundamental problems remain unsolved.  
  
The paper is organized as follows: we start by introducing the microscopic model we are interested in. The following sections are devoted to the rather long computation which starts from the microscopic model and ends with the 
entropy functional describing the spatial density in the system. The main steps of the computation are outlined at the beginning of the third section.
Finally, we draw some physical conclusions from the derived coarse-grained equations. 
%% entropy functional in section \ref{sec:physics}.

\section{Model}
\label{sec:models}
We consider a system of $N$ active particles, $i=1,\ldots,N$, moving in a two-dimensional space. 
The position of the i-th particle is given by $\mathbf{x}_i=(x_i,y_i)$ and what we refer to as its active velocity (AV) by $v(n_i) \mathbf{u}(\theta_i)$, 
where  $\mathbf{u}(\theta_i)\equiv (\cos(\theta_i), \sin(\theta_i))$ defines  the direction of the AV and $v(n_i) \equiv v_0 \tilde{v}(n_i)$ its norm, with $v_0$ a constant and $\tilde{v}(n_i)$ a function that depends on $n_i$.    
The term $n_i$ refers to the local density around the i-th particle. 
More specifically,  $n_i = \sum_{j=1}^N g(|\mathbf{x}_i-\mathbf{x}_j|/R)$ where  the function $g(\Delta/R)$, with $\Delta = |\mathbf{x}_i-\mathbf{x}_j|$ defines the interaction range.  
Finally, we consider an over-damped dynamics for the evolution of the $\mathbf{x}_i=(x_i,y_i)$ and $\theta_i$ such that the equations of motion of the i-th particle take the form:
\begin{eqnarray}
\label{eq:movx} 
\dot{\mathbf{x}}_i &=& v(n_i) \mathbf{u}(\theta_i) + \sqrt{2D_x}  \vec{\sigma}_i(t) \\
\dot{\theta}_i &=& - \frac{\gamma}{n_i}\sum_{j=1}^N g(|\mathbf{x}_i-\mathbf{x}_j|/R)\sin(\theta_i-\theta_j) +\sqrt{2D_\theta} \eta_i(t) \, .
\label{eq:mova}
\end{eqnarray}
$\vec{\sigma}_i(t) = (\sigma^x_{i}(t),\sigma^y_{i}(t))$ and $\eta_i(t)$ for $i=1,\ldots,N$ are white, gaussian and uncorrelated noises with unit covariance. These noises represents a ``bath" with very short time 
correlations or memory. The alignment sensitivity $\gamma$ is not directly related to the fluctuation amplitude $D_\theta$ through an Einstein relation. A suitable option for the function $g$ is to take $g(\Delta/R)=1$ for $\Delta/R \leq 1$ and $0$ otherwise, definition by which $R$ defines the interaction range. 
Notice that the model definition, in particular $v(n_i)$, implies a mesoscale description; i.e., we assume that there is a  microscopic physical mechanism that leads to $v(n_i)$. 
Furthermore, we require $v(x)$ to be a differentiable function, which is not necessary applicable to lattice models with strict exclusion rules~\cite{soto2014}.

It is convenient to write the equations of motion in adimensional form. Calling $L$ the box size, we write $x_i=L\tilde{x}_i$, $y_i=L\tilde{y}_i$.
We also rescale the time $t=\tilde{t}/D_\theta$, and adopt $v_0$ as the velocity scale. By using $\vec{\sigma}_i( \widetilde{t}/D_{\theta})=
\sqrt{D_{\theta}}\vec{\sigma}_i(\widetilde{t})$ and $\eta_i({\widetilde{t}/D_{\theta}})=\sqrt{D_{\theta}}\eta_i({\widetilde{t}})$, we arrive to: 
\begin{eqnarray}
\frac{d\mathbf{\tilde{x}}_i}{d\tilde{t}} &=& \epsilon \tilde{v}(n_i) \mathbf{u}(\theta_i) + \epsilon \sqrt{2 \tilde{D}_x}  \vec{\sigma}_i(\tilde{t}) \\
% \frac{d\tilde{x}_i}{d\tilde{t}} &=&  \varepsilon \tilde{v}_i\cos \theta_i \\
% \frac{d\tilde{y}_i}{d\tilde{t}} &=&  \varepsilon \tilde{v}_i\sin \theta_i  \\
\frac{d\theta_i}{d\tilde{t}}&=& - \frac{\bar{\gamma}}{n_i} \sum_{j=1}^N g((\tilde{\mathbf{x}}_i-\tilde{\mathbf{x}}_j)/\alpha)\sin(\theta_i-\theta_j) + \sqrt{2}~\eta_i(\tilde{t}) \, ,
\end{eqnarray} 
where we have introduced the dimensionless parameters $\bar{\gamma}=\gamma/D_\theta$, $\alpha=R/L$, $\varepsilon=v_0/(LD_\theta)$, and $\tilde{D}_x = D_x D_{\theta}/v_0^2$, 
and so  $n_i =  \sum_{j=1}^N g((\tilde{\mathbf{x}}_i-\tilde{\mathbf{x}}_j)/\alpha)$. 
We notice that  $\tilde{D}_x = D_x D_{\theta}/v_0^2$ is the ratio between the passive ($D_x$) and active ($v_0^2/[2 D_{\theta}$]) diffusion coefficient, which we consider to be of order $\varepsilon^0$.  From now on, we  work with the adimensional equations and drop the $\tilde{}$ to simplify the notation. In the following, we will assume $\alpha\to 0$.

\section{Main computation}
\label{sec:main}
As is clear from the scaling introduced through the parameter $\varepsilon$, we are interested in the situation where the dynamics over the angles 
$\theta_i$ is fast with respect to the spatial dynamics. Furthermore, we want to study the density of particles and their local mean orientation over
large length scales. Our ultimate goal is to obtain a static large deviation principle yielding an entropy functional, which describes the fluctuations 
of the spatial empirical density, along the lines of \cite{tailleur2008,tailleur2013}, but accounting for the existence of a velocity alignment mechanism.
To make the forthcoming computations easier to follow, we outline here the general scheme:
\begin{enumerate} 
\item Write an effective equation for the phase space empirical density (that is in the variables $x,y$ and $\theta$), keeping the finite $N$ fluctuations. This leads to a stochastic PDE with a noise term of order $1/\sqrt{N}$. 
\item Take advantage of the time-scale separation to obtain a closed effective dynamics for the empirical density of the slow spatial variables $x,y$. 
\item Write a functional Fokker-Planck equation for the  macroscopic density  field $\rho$.
%% : $\rho$ in the disordered case; there would be $\rho$ and the local order parameter in the ordered phase.
\item Look for the stationary probability density of the macroscopic field $\rho$,  under the asymptotic form 
$e^{NS[\rho]}$ and solve for $S$ at leading order in $N$. 
%\item Write a dynamical large deviation principle for the spatial empirical density, similar to those obtained in \cite{DawsonGartner}, or
%assumed in Eq.(1.7) of \cite{MFT02}.
%\item Obtain from this dynamical large deviation principle the Hamilton-Jacobi functional equation satisfied by the entropy functional we are looking for.
%\item Solve this Hamilton-Jacobi functional equation.
\end{enumerate}
% Note that this last step amounts to solve a functional Hamilton-Jacobi equation, as shown in \cite{MFT02}, which in general is not possible explicitly; it will turn out to be possible in the disordered phase. Beyond the stationary large deviation principle, we can also write a dynamical large deviation principle for the spatial empirical density, similar to those obtained in \cite{DawsonGartner}, or assumed in Eq.(1.7) of \cite{MFT02}. 

\subsection{Dynamical equation for the phase space empirical density}

We assume that $N$ is large, but finite, and denote by $f_d(\mathbf{x},\theta,t)$ the empirical density of particles in the 3D space given but $[x,y,\theta]$. 
The temporal evolution of $f_d(\mathbf{x},\theta,t)$ is expressed in terms of the following stochastic partial differential equation: 
\begin{eqnarray}
\frac{\partial f_d}{\partial t}&=& - \varepsilon \nabla . \left(v(\rho(\mathbf{x},t))) \mathbf{u}(\theta) f_d(\mathbf{x}, \theta,t) \right) 
 +\frac{\bar{\gamma}}{\rho(\mathbf{x},t)} \frac{\partial }{\partial \theta} \left(f_d(\mathbf{x},\theta,t)  \int d\theta' \text{sin}(\theta - \theta')f_d(\mathbf{x},\theta',t) \right)  \nonumber \\
&&+ \frac{\partial^2 f_d}{\partial \theta^2}+ \varepsilon^2 D_x \nabla^2 f_d +\sqrt{\frac{2}{N}}\frac{\partial }{\partial \theta}\left( \eta(\mathbf{x},\theta,t) \sqrt{f_d}\right) \nonumber \\ 
&&+\varepsilon \frac{\sqrt{2D_x}}{\sqrt{N}} \nabla\cdot \left( \vec{\sigma}(\mathbf{x},\theta,t) \sqrt{f_d}\right) \, ,
\label{eq:R0}
\end{eqnarray}
where $\eta(\mathbf{x},\theta,t)$ and $\vec{\sigma}(\mathbf{x},\theta,t)$  are gaussian noises, delta-correlated in time and space, and $\rho(\mathbf{x},t)$ is the empirical spatial density defined by:
\begin{eqnarray}
 \rho(\mathbf{x},t) = \int d\theta f_d(\mathbf{x},\theta,t).
 \end{eqnarray} 
This can be shown at a  formal level by following~\cite{Dean96}. 
% or via  large-deviation analysis as in~\cite{Dawson}.

\subsection{Averaging step - Time scale separation}

We now take Fourier components of $f_d$ of increasing order, stopping the expansion as soon as possible. This is a standard strategy, see e.g. \cite{bertin2009}.  Notice  that finite $N$ fluctuations are taken into account. We use the following notations: $\mathbf{P}=(P_x,P_y)$, with
\begin{eqnarray}
\mathbf{P}(\mathbf{x},t) &=& \int d\theta \mathbf{u}(\theta) f_d(x,y,\theta,t) \, .
\end{eqnarray}
Integrating Eq. (\ref{eq:R0}) respectively over $d\theta, \cos \theta d\theta$ and $\sin \theta d\theta$, we obtain: 
\begin{eqnarray}
\frac{\partial\rho}{\partial t} &=&  - \varepsilon \nabla.(v \vec{P}) + \varepsilon^2 D_x \nabla^2 \rho  + \varepsilon \frac{\sqrt{2D_x}}{\sqrt{N}} \nabla\cdot \left( \vec{\xi}(x,y,t) \right)\label{eq:rho}\\ 
\frac{\partial \mathbf{P}}{\partial t} &=& - \frac{1}{2}\varepsilon \nabla \left( v \rho \right)  + \left( \frac{\bar{\gamma}}{2}-1 \right) \mathbf{P} +\varepsilon^2 D_x \nabla^2 \mathbf{P}  + \sqrt{\frac{2}{N}} \vec{\eta}(x,y,t) + O(\frac{\varepsilon}{\sqrt{N}}) 
\label{eq:Py}
\end{eqnarray}
where the noises are defined by
\begin{eqnarray}
\eta_x(x,y,t) &=& \int d\theta \sin \theta \sqrt{f_d}~\eta(x,y,\theta,t) \\
\eta_y(x,y,t) &=& -\int d\theta  \cos \theta \sqrt{f_d}~\eta(x,y,\theta,t)  \\
\vec{\xi}(x,y,t) &=& \int  d\theta \sqrt{f_d}~\vec{\sigma}(x,y,\theta,t) \,.
\end{eqnarray}
By construction, the noises are gaussian, delta-correlated in time and space. Furthermore
\begin{eqnarray}
\langle \eta_x (x,y,t) \eta_x (x',y',t') \rangle &\simeq&  \delta(x-x')\delta(y-y')\delta(t-t')\frac12 \rho(x,y,t) \\
\langle \eta_y (x,y,t)  \eta_y (x',y',t') \rangle  &\simeq&  
 \delta(x-x')\delta(y-y')\delta(t-t')\frac12 \rho(x,y,t)  \\ 
 \langle \eta_x (x,y,t) \eta_y (x',y',t') \rangle  &\simeq&  0 \\
 \langle \xi(x,y,t) \xi(x',y',t') \rangle \rangle  &=& 
 \delta(x-x')\delta(y-y')\delta(t-t') \rho(x,y,t) \, .
\end{eqnarray}
Consistently with our approximation, we have dropped in the noise correlation all Fourier coefficients beyond the first. 
Notice that in Eq.~(\ref{eq:Py}) we have neglected higher order Fourier coefficients  since we  are interested in characterizing the system dynamics in the disordered phase -- i.e. when collective motion is not observed. This implies that our approximation is only valid below the onset of collective motion. 
Moreover, without higher Fourier components, Eq.~(\ref{eq:Py}) predicts that  
 $|\mathbf{P}|$ grows unboundedly  for $\bar{\gamma}>2$. To obtain a system of equations that is physically well-behaved in the ordered phase, we have to go at least one component further in Fourier. Such extra Fourier component is connected to the nematic order, while $\mathbf{P}$ to polar order. To understand the disordered phase, which is our objective here, we insist that it is enough to develop up to polar order. 
Furthermore, we stress that  Eq.~(\ref{eq:Py}) is consistent with a small   $\varepsilon$ expansion.

%\frac{\partial P_y}{\partial t} &=& - \frac12 \varepsilon \frac{\partial}{\partial y}\left( v \rho \right)
%+\left( \frac{\bar{\gamma}}{2}-1 \right)P_y +\sqrt{\frac{1}{N}} \eta_y(x,y,t)
%\label{eq:Py}

If $\bar{\gamma}$ smaller than 2 and not too close to 2, $\mathbf{P}$ very quickly reaches its stationary value and remains small: particle motion is locally 
disordered, since the interaction promoting alignment 
is not strong enough to create a local orientational order. In this regime, we can take the l.h.s. of Eq.~(\ref{eq:Py}) to be $0$ in order to determine the stationary value of $\mathbf{P}$. Neglecting terms of order $\varepsilon^2$ and $\varepsilon/\sqrt{N}$, we obtain: 
\begin{eqnarray}
%% P_x &=& \varepsilon \frac{1}{2(1-\frac{\bar{\gamma}}{2})} \partial_x(v\rho) +\sqrt{\frac{1}{N}} \frac{1}{1-\frac{\bar{\gamma}}{2}} \eta_x\\
\mathbf{P} &=& \varepsilon \frac{-1}{2(1-\frac{\bar{\gamma}}{2})} \nabla(v(\rho)\rho) +\sqrt{\frac{2}{N}} \frac{1}{1-\frac{\bar{\gamma}}{2}} \vec{\eta}
 \label{eq:P_only}
%% P_x &=& \varepsilon \frac{1}{2(1-\frac{\bar{\gamma}}{2})} \partial_y(v\rho) +\sqrt{\frac{1}{N}} \frac{1}{1-\frac{\bar{\gamma}}{2}} \eta_y
 \end{eqnarray}
These computations are formal, and could in general lead to incorrect results: one should in particular be cautious about the meaning of the noise term, which is multiplicative. However, since we will eventually take a small noise limit (large $N$), this formal approach will correctly yield the leading order in $N$. We insert Eq.(\ref{eq:P_only}) into (\ref{eq:rho}) and look for the 
long-time behavior of $\rho$  by introducing a new time-scale $\tilde{t}=\varepsilon^2 t$:  
\begin{eqnarray}
\frac{\partial\rho}{\partial t} &=& \frac12  \nabla \cdot\left( \frac{v}{1-\frac{\bar{\gamma}}{2}}\nabla[v(\rho)\rho]\right)
+ D_x \nabla^2 \rho \nonumber \\
&& + \frac{\sqrt{2D_x}}{\sqrt{N}} \nabla\cdot \left( \vec{\xi}(\vec{r},t) \right)+ \sqrt{\frac{2}{N}}\nabla \cdot\left( \frac{v}{1-\frac{\bar{\gamma}}{2}}\vec{\eta}\right) \, ,
\label{eq:rho_final}
\end{eqnarray}
where again we have dropped the $\tilde{~}$ and replace  $\vec{\eta}$ by $-\vec{\eta}$. Notice that due to the involved change of time-scale, all $\epsilon$'s have disappeared of the final equation, and both noise terms give a contribution. 
%each noise term contributed with $\epsilon$. 
The expansion in powers of $\varepsilon$ is formally consistent, in the sense that further Fourier components would contribute terms which are formally of higher order. This means that one can hope that Eq.~(\ref{eq:rho_final}) is in some sense exact in the limit $N\to \infty~,~\varepsilon \to 0$. 

Eq.~(\ref{eq:rho_final}) can be expressed in a more compact notation in the following way: 
\begin{eqnarray}
\frac{\partial\rho}{\partial t} &=&	U[\rho](\vec{x}) +\frac{1}{\sqrt{N}} \nu(\vec{x},t)
\label{eq:rho2}
\end{eqnarray}
where 
\begin{equation}
U[\rho](\vec{x}) = \frac12 \nabla \cdot\left( \frac{v(\rho)}{1-\frac{\bar{\gamma}}{2}}\nabla[v(\rho)\rho] \right) +D_x \nabla^2 \rho  
\label{eq:U}
\end{equation}
and $\langle \nu(x,y,t)\nu(x',y',t')  \rangle=D[\rho](\vec{x},\vec{x}')\delta(t-t')$ with
\begin{eqnarray}
D[\rho](\vec{x},\vec{x}')&=&  \partial_x\partial_{x'}[b[\rho](\vec{x}) \delta(\vec{x}-\vec{x}')] +\partial_y\partial_{y'}[b[\rho](\vec{x}) \delta(\vec{x}-\vec{x}')] 
\label{eq:noise}
\end{eqnarray}
where
\[
b[\rho] = 2D_x\rho+ \frac{\rho v^2(\rho)}{(1-\frac{\bar{\gamma}}{2})^2}
\]
We have combined here the two independent gaussian noise terms into a single one.

\subsection{Functional Fokker-Planck equation}

From Eq.~(\ref{eq:rho2}), one can write a functional Fokker-Planck equation (see for details~\cite{gardiner}) for the probability distribution of the density field $\mu_{t}\left[\rho\right]$:
\begin{eqnarray}
\nonumber \frac{\partial \mu_{t}}{\partial t} &=& -\int d\vec{x} \frac{\delta }{\delta \rho(\vec{x})}\left( U[\rho](\vec{x}) \mu_{t} \right) \\
 &+& \frac{1}{2N}\int d\vec{x} \frac{\delta}{\delta \rho(\vec{x})}\left\{ \int d\vec{x}' D[\rho](\vec{x},\vec{x}')\frac{\delta}{\delta \rho(\vec{x}')} \mu_{t} \right\}
\label{eq:funcFP}
\end{eqnarray}
%
%% Introduce definition of the empirical density.
%
Note that we have here assumed an interpretation of the noise $\nu$ corresponding to Ito's convention. This has no consequence at leading order in $N$. 
We look for a stationary solution taking the asymptotic form
\begin{equation}
\label{eq:prob_rho}
\mu\left[\rho\right]\sim e^{N S[\rho]}
\end{equation}
and compute $S$ at leading order in $N$. The drift $U$ and the noise correlation $D$ depend on $\rho$. However, we see that the relevant terms at leading order in $N$ are obtained when the functional derivatives with respect to $\rho$ act on $\mu_t$ rather than on $U$ or $D$.
This leads to the following equation for $S$:
\begin{eqnarray}
  \int U[\rho](\vec{x})\frac{\delta S}{\delta\rho(\vec{x})}d\vec{x} &=& 
  \frac12 \left\{\int d\vec{x} \left[\partial_x \frac{\delta S}{\delta\rho(\vec{x})} 
\int d\vec{x}' \partial_{x'}  \left(b[\rho](\vec{x'})\delta(\vec{x}-\vec{x}')\right)
\frac{\delta S}{\delta\rho(\vec{x}')}\right]
  \right.  \nonumber\\ 
  && + \left. \int d\vec{x} \left[\partial_y \frac{\delta S}{\delta\rho(\vec{x})}
\int d\vec{x}' \partial_{y'}  \left(b[\rho](\vec{x}') \delta(\vec{x}-\vec{x}')\right)
\frac{\delta S}{\delta\rho(\vec{x}')}\right] \right\}  \nonumber \\
  &=& -\frac12 \int d\vec{x}  \nabla\cdot \left(b[\rho](\vec{x}) \nabla \frac{\delta S}{\delta\rho(\vec{x})}\right)
  \frac{\delta S}{\delta\rho(\vec{x})} 
\end{eqnarray}
By comparison and using the expression (\ref{eq:U}) for $U$, one sees that a sufficient condition to find $S$ is to solve the equation
\begin{equation}
\frac12 b[\rho] \nabla \frac{\delta S}{\delta\rho(\vec{x})} = -\frac12 \frac{v(\rho)}{1-\frac{\bar{\gamma}}{2}} \nabla(\rho v(\rho)) -D_x\nabla \rho
\end{equation}
Formally, it can be shown that this expression represents the equilibrium condition for $\mu\left[\rho\right]$, that is the condition for a zero-flux solution of the functional Fokker-Planck equation, Eq. (\ref{eq:funcFP}). 
Equivalently, this corresponds to the reversibility of the dynamics given by Eq. (\ref{eq:rho2})  with respect to the density  $\mu[\rho]$. 
We look for a solution $S$, which is a local functional of $\rho$, and of the form:
\begin{equation}
S[\rho] = \int d\vec{x} s(\rho(\vec{x})) \label{eq:entrop}
\end{equation}
where the entropy density $s$ is a real function to be determined. One finds
\begin{eqnarray}
s"(\rho) &=& - \left( \frac{v^2(\rho)+\rho v(\rho) v'(\rho)}{\left(1 - \frac{\bar{\gamma}}{2}\right)b[\rho]} + \frac{2D_x}{b[\rho]} \right) \label{eq:ddentrop}
% \\
%&=& -\frac{v^2+\rho vv'}{2D_x(1-\frac{\bar{\gamma}}{2})\rho+\frac{\rho v^2}{1-\frac{\bar{\gamma}}{2}}}-\frac{2D_x\rho}{2D_x\rho+\frac{\rho v^2}{\left(1-\frac{\bar{\gamma}}{2}\right)^2}} \nonumber
\end{eqnarray}
When the integration of Eq.~(\ref{eq:ddentrop}) is possible, the function $s$ can be explicitly retrieved. 
This is the main result of this article. In the absence of alignment, i.e. 
 $\bar{\gamma}=0$, Eq.~(\ref{eq:ddentrop}) leads to the same results derived in~\cite{tailleur2013}. For instance, it is straightforward to see that if in addition we make $D_x=0$, 
the spinodal line is given by the condition $dv/d\rho=-v/\rho$ as explained in~\cite{tailleur2008,tailleur2013}.

Notice that the procedure followed to arrive to the free energy, consisted in deriving a stochastic equation for the 
empirical density, and using  Ito's calculus to obtain an expression for the density field $\mu$, in turn connected to the free energy $S$. 
We stress that the described procedure is fundamentally different from recent approaches~\cite{speck2013} used to describe phase separation 
in non-aligning active particles, where a free energy is obtained by deriving first a non-fluctuating BBGKY hierarchy of equations, performing a perturbation expansion, 
and making a direct analogy between the derived equation for the density field at second leading order  and the Cahn-Hilliard equation, whose free energy is well known. 
Our derivation, on the contrary, contains finite $N$ fluctuations, going beyond mean-field, and allows us to obtain directly a free-energy-like functional form, which 
is not necessary a Cahn-Hilliard free energy with a local cubic term as in~\cite{speck2013}. 
In the following section, we discuss the physical meaning of equations here derived. 
 
\section{Physical discussion and final remarks}
\label{sec:physics}

%%% Fig 1 %%%%%%%%%%%%%%
\begin{figure}
\centering
\resizebox{\columnwidth}{!}{\rotatebox{0}{\includegraphics{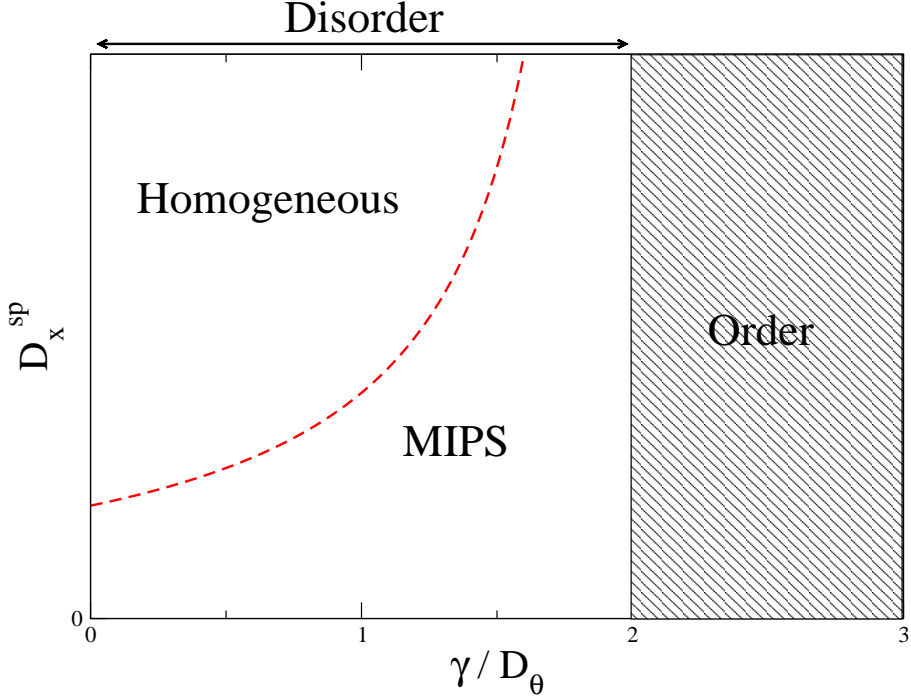}}}
\caption{Sketch of phase diagram: $D^{SP}_x$
  vs. $\bar{\gamma}=\gamma/D_{\theta}$ at fixed density. The figure
  indicates that the transition between the (orientationally)
  disordered and ordered phase is controlled by the ratio
  $\bar{\gamma} = \gamma/D_{\theta}$.  In the disordered phase, the
  red dashed curve represents the spinodal line that diverges
  (according to our approximation) as the onset of 
  orientational order is approached. As we come close to the
  disorder-order transition density fluctuations also diverge (see
  text).  Above the red curve there is only one homogeneous phase,
  while below the system phase separates, i.e., the homogeneous phase
  is no longer stable.
It has been shown that the ordered phase can exhibit a zoo of patterns, see~\cite{peruani2011,tailleur2012}. 
In the sketch we have assumed that $v(\rho)=\exp(-\lambda \rho)$, with $\lambda$ large enough to allow MIPS. 
Notice that in absence of alignment interactions, $D_x^{sp}$ is a constant, whose value corresponds to $\bar{\gamma}=\gamma/D_{\theta}=0$ in the figure. 
We remind the reader that $D_x^{sp}$ is a dimensionless parameter, which means that for non-aligning particles the spinodal is given by $D_x D_{\theta} / v_0^2=D_x^{sp}(\bar{\gamma}=0)$.  
%%% The dot-dashed blue line corresponds to the spinodal scenario in absence of alignment interactions. 
%
} \label{fig:PhaseDiagram}
\end{figure}
%%% Fig 1 %%%%%%%%%%%%%%

Let us review the results we obtained. 
The equations of motion~(\ref{eq:movx}) and~(\ref{eq:mova}) were our starting point. 
We required $v(x)$ to be a differentiable function, which implies that our derivation is, in principle, not adequate to describe sharp interfaces as the ones observed in lattice models with strict exclusion rules~\cite{soto2014}. 
Under these assumptions, we derived an equation for the empirical density, Eq.~(\ref{eq:R0}), following~\cite{Dean96}. 
This equation becomes an exact description only in the limit of infinite $N$ and infinite densities.  
Thus, for finite but large $N$, Eq.~(\ref{eq:R0}) should provide a good  description for the particle density of a system whose  microscopic dynamics is given by Eqs. (\ref{eq:movx}) and (\ref{eq:mova}).  
Our goal has been to derive an entropy-like functional for the particle density. 
In order to do that, we made an expansion in Fourier of Eq.~(\ref{eq:R0}), given by Eqs.~(\ref{eq:rho}) and~(\ref{eq:Py}), up to polar order $\mathbf{P}$, and made use of the fast relaxation of $\mathbf{P}$ with respect of the temporal evolution of $\rho$, 
to express $\mathbf{P}$ as function of  $\rho$ and its gradients. 
Such expansion up to polar order, as well as the  separation of time-scales between the temporal evolution of $\mathbf{P}$ and $\rho$, are exclusively valid in the disordered phase. 
%
%
% evolution of $\rho$
%
%we expressed the polar order $\mathbf{P}$ as function of the density $\rho$ and its gradients. 
%%
%We made use of the fast relaxation of $\mathbf{P}$ with respect of the temporal evolution of $\rho$. 
%%
%Such a separation of time-scales played a crucial role in the derivation of the entropy functional. 
%
% Strictly speaking, this is true in the disordered phase. 
%
From Eq.~(\ref{eq:Py}), we can easily see that  the onset of local (orientational) order occurs  for $\bar{\gamma}/2-1>0$, that is, 
when the angular diffusion $D_{\theta}$ is such that $D_{\theta}<\gamma/2$. 
This means that the entropy functional, given by  Eqs. \eqref{eq:entrop} and \eqref{eq:ddentrop}, is valid for $D_{\theta}>\gamma/2$,
see solid vertical line in Fig.~\ref{fig:PhaseDiagram}.

%%% Fig 1 %%%%%%%%%%%%%%
\begin{figure}
\centering
\resizebox{\columnwidth}{!}{\rotatebox{0}{\includegraphics{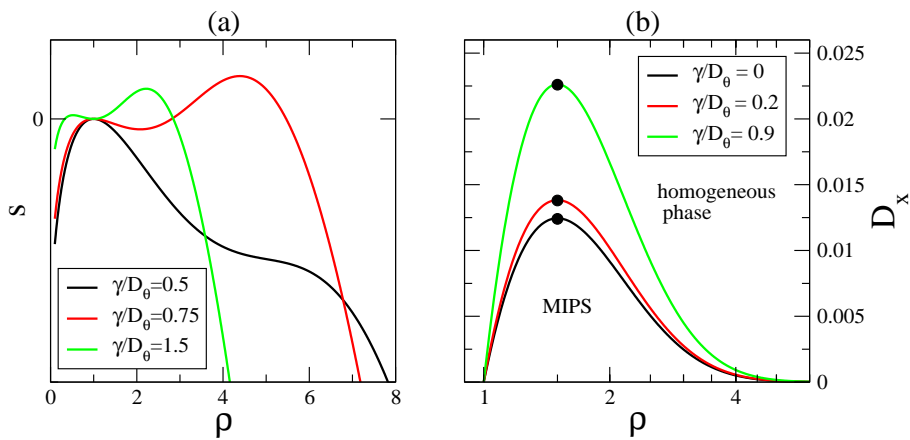}}}
\caption{Phase separation in the disordered phase is controlled by the interplay between $\bar{\gamma}=\gamma/D_{\theta}$ and $D_x$. (a) Entropy $s(\rho)$, derived from Eq.~(\ref{eq:ddentrop}), 
for various values of $\bar{\gamma}=\gamma/D_{\theta}$. (b) Phase diagram $D_x$ vs. $\rho$ for various values of $\bar{\gamma}=\gamma/D_{\theta}$, see Eq.~(\ref{eq:spinodal}). While below each curve 
$D^{SP}_x(\rho)$, the system is motility-induced phase separated (MIPS), above $D^{SP}_x(\rho)$ the system remains homogeneous (between the binodal and spinodal lines, the homogeneous phase is only metastable). This phase diagram is the counterpart of the classical gas-liquid phase diagram $T-\rho$. 
The critical point, defined by ($\rho_c$, $D_x^{crit}$), see black dots and text, marks the value of $D_x$ above which there is no more phase transition.   
In the figure we have assumed that $v(\rho)=\exp(-\lambda \rho)$. The curves in (a) and (b) correspond  to  $\lambda=1$.  
%%% The dot-dashed blue line corresponds to the spinodal scenario in absence of alignment interactions. 
%
} \label{fig:MIPS}
\end{figure}
%%% Fig 1 %%%%%%%%%%%%%%

To understand the physical meaning of the derived equations, let us
adopt a concrete functional form for $v(\rho)$, e.g. $v(\rho)=
\exp(-\lambda \rho)$.  Notice that the qualitative features discussed
  below do not depend on the precise functional form for $v(\rho)$,
  provided it is decreasing. 
Phase separation occurs below the binodal line defined by the double tangent construction on the $s(\rho)$ curve. Easier to detect and compute is the spinodal line, below which an homogeneous phase cannot be stable. 
The spinodal can be found solving $s''(\rho)=0$, using expression Eq. \eqref{eq:ddentrop}. One finds
\begin{equation}
\frac{e^{-2\lambda \rho}(\lambda \rho -1)}{2-\bar{\gamma}} = D_x^{sp}(\rho)
\label{eq:spinodal}
\end{equation}
For $\rho,\lambda$ such that $\lambda\rho>1$, Eq. \eqref{eq:spinodal} is represented by the red dashed line in Fig.~\ref{fig:PhaseDiagram}. Above this line, the homogeneous solution is stable; below it, spinodal decomposition (Motility-Induced Phase Separation) occurs.
Notice that this line, which corresponds to a dynamical instability, can also be obtained by linearizing Eq.~(\ref{eq:rho2}), without noise, around a spatially homogeneous solution. Eq.~\eqref{eq:ddentrop} also gives access to the metastable 
regions around the spinodal line.
%
%To find the spinodal line, we can study the stability of an spatial homogeneous% solution of the form $\rho(\mathbf{x},t)=\rho_0$ with $\rho_0$ a constant. 
%
%Let us assume that $\rho(\mathbf{x},t)=\rho_0 + \delta_{\rho}(\mathbf{x},t)$,  $N \to \infty$, and insert this expression into  Eq.~(\ref{eq:rho2}). 
%
%Keeping linear terms in $\delta_{\rho}$, we obtain the linear stability of the homogeneous solution, which is indicated by the red dashed-line in Fig.~\ref{fig:PhaseDiagram}. 
%
Figure~\ref{fig:PhaseDiagram} shows that the critical spatial
diffusion $D_x^{sp}$ below which the homogeneous solution is unstable
strongly depends on $\bar{\gamma} =\gamma/D_{\theta}$: the divergence of
$D_x^{sp}$ in our approximation is connected to the term in $1/(1-\bar{\gamma}/ 2)$ in
Eq.~(\ref{eq:U}).  This high sensitivity of the spinodal line to the
alignment strength, i.e. $\gamma$, is consistent with results obtained in
simulations~\cite{peruani2011}.
As expected, the entropy $s(\rho)$ is also affected by $\bar{\gamma}$ as shown in Fig.~\ref{fig:MIPS}(a). 

Thus, our results indicate that MIPS as described in~\cite{tailleur2008,tailleur2013} also occurs in the presence of alignment interactions in the disordered phase, with $\bar{\gamma}$  affecting the spinodal (as well as 
binodal) line as shown in Fig.~\ref{fig:MIPS}(b). Notice that Eq.~(\ref{eq:spinodal}) allows us to draw the phase diagram $D_x-\rho$, which is the  counterpart of the classical gas-liquid phase diagram $T-\rho$. 
From this expression we can obtain the so-called critical point ($\rho_c$, $D_x^{crit}$), with $\rho_c = 3/(2\lambda)$ and $D_x^{crit}\simeq 0.0249/(2-\bar{\gamma})$ for the chosen functional form of $v(\rho)$. 
For $D_x>D_x^{crit}(\bar{\gamma})$, Eq.~\eqref{eq:spinodal} has 
no solution, which implies that in this case, the homogeneous phase is stable for any density 
and value of $\lambda$. In short, MIPS has disappeared above  $D_x^{crit}(\bar{\gamma})$. Hence, the critical 
value of the spatial diffusion needed to destroy the MIPS strongly increases 
when the alignment interactions increase. Moreover, in the current approximation, it diverges as order-disorder transition is approached, which suggests that there is no critical point in the ordered phase. 

Finally, we stress  that at this level of approximation, $D_x$ (i.e., the original, dimensional, spatial diffusion constant)  does not affect the orientational order transition point (more precisely the instability of the homogeneous disordered phase), while $\gamma$ and $D_{\theta}$ play a role on both, the disordered and ordered phase. 
Now, let us turn to the analysis of density fluctuations related to the noise term present in Eq.~(\ref{eq:rho2}). From Eq.~(\ref{eq:ddentrop}), we learn that in the homogeneous phase, an alignment interaction 
$\gamma>0$ makes $s''$ smaller in absolute value, and thus it increases the density 
fluctuations.  
% , the homogeneous and MIPS phase.  
%
Notice that density fluctuations diverge for $s'' \to 0$. This occurs on the spinodal line, as well as in the MIPS phase, as we approach 
the instability of the (homogeneous) disordered phase, i.e. $\gamma \to 2 D_{\theta}$. 
It is worth noticing that for $D_x = 0$, $\gamma$ has no influence on the phase diagram shown in Fig.~\ref{fig:PhaseDiagram}, while it still has 
on the density fluctuations. 
A note of caution is in order here. According to the proposed approach, $D_x^{sp}$, $D_x^{crit}$, as well as density fluctuations diverge as the disorder-order transition is approached. As explained above, our approximation is not valid in the limit of   $\gamma \to 2 D_{\theta}$. 
While we can be sure that $D_x^{sp}$, $D_x^{crit}$ and density fluctuation increase as we approach the onset of collective motion, we cannot ensure that the system behavior at the 
disorder-order transition  or in its vicinity is as predicted by the present approach.  
 
In summary, our calculations indicate that phase separation can occur 
in the disorder phase with the alignment strength -- more specifically with the ``distance" to the instability of the (homogeneous) disordered phase -- controlling the position of the 
spinodal line involved in the MIPS as well as the size of density fluctuations. 
In short, we have generalized the approach of~\cite{tailleur2008,tailleur2013} in order to account for the presence of a velocity alignment
mechanism.  While the nature of the described phase separation remains
a MIPS as observed in non-aligning systems~\cite{fily2012,redner2013,fily2014,valeriani2013,wysocki2013,cates2013,speck2013}, this does not exclude that in the
orientationally ordered phase, phase separation can be of a
different nature as suggested in~\cite{peruani2013}.

%\julien{\it Another effect we have not discussed: at very high density, $D_x$ dominates so that there is no MIPS. This makes possible the following. 
%Fix everything, and
%increase density; then it is possible that the system is first homogeneous (low density), then crosses the spinodal and phase separates, then crosses again the spinodal, and becomes homogeneous again (high density). Imagine we control also 
%$\gamma$ (or $D_x$, or $\lambda$). Then we may first decrease $\gamma$ such that no MIPS is possible. then increase density from low to high, then increase $\gamma$ again to its original value. We have followed another path in phase space to go from low density homogeneous to high density homogeneous, but this time without phase separation. In other words, we have gone around a critical point. This is not specific to the case with alignment however; I don't know if it is interesting... }

{\bf acknowledgements} We acknowledge enlightening discussions with
O. Dauchot, P. Degond, and J. Tailleur and financial support from the
PEPS-PTI ``Anomalous fluctuations in the collective motion of
self-propelled particles".
The suggestion by J. Tailleur of adding a white noise to
Eq.(\ref{eq:movx}) has proved to be very fruitful  
for the present study. 
FP thanks the Kavli Institute for Theoretical Physics (University of California, Santa Barbara) and the organizers of the bioacter14 program for hospitality and financial support.


\begin{thebibliography}{99}
%\expandafter\ifx\csname natexlab\endcsname\relax\def\natexlab#1{#1}\fi
%\expandafter\ifx\csname bibnamefont\endcsname\relax
%  \def\bibnamefont#1{#1}\fi
%\expandafter\ifx\csname bibfnamefont\endcsname\relax
%  \def\bibfnamefont#1{#1}\fi
%\expandafter\ifx\csname citenamefont\endcsname\relax
%  \def\citenamefont#1{#1}\fi
%\expandafter\ifx\csname url\endcsname\relax
%  \def\url#1{\texttt{#1}}\fi
%\expandafter\ifx\csname urlprefix\endcsname\relax\def\urlprefix{URL }\fi
%\providecommand{\bibinfo}[2]{#2}
%\providecommand{\eprint}[2][]{\url{#2}}


%% \bibitem{humans} D. Helbing, I. Farkas, and T. Vicsek, Nature (London) {\bf 407}, 487 (2000).

%% Birds
\bibitem{cavagna10}  Cavagna, A. et al.: Scale-free correlations in starling flocks. Proc. Natl. Acad. Sci. {\bf 107}, 11865-11870 (2010).
%% Cavagna, A. et al.: Scale-free correlations in starling flocks. Proc. Natl. Acad. Sci. {\bf 107}, 11865--11870 (2010).

\bibitem{bhattacharya10} Bhattacharya, K., Vicsek, T.: Collective decision making in cohesive flocks. New J. Phys. {\bf 12}, 093019 (2010).
%% Bhattacharya, K.,  Vicsek, T.: Collective decision making in cohesive flocks. New J. Phys. {\bf 12}, 093019 (2010). 
%%

\bibitem{buhl06}  Buhl, J. et al.: From disorder to order in marching locusts. Science {\bf 312}, 1402-1406 (2006).
%% J. Buhl et al.,  Science {\bf 312}, 1402 (2006).

\bibitem{romanczuk09} Romanczuk, P., Couzin, I.D., and Schimansky-Geier, L.: Collective motion due to individual escape and pursuit response. Phys. Rev. Lett. {\bf 102}, 010602 (2009).

%% P. Romanczuk, I.D. Couzin, and L. Schimansky-Geier,  Phy. Rev. Lett. {\bf 102}, 010602 (2009).

\bibitem{zhang10} Zhang, H.P. et al.: Collective motion and density fluctuations in bacterial colonies. Proc. Natl. Acad. Sci. {\bf 107}, 13626-13630 (2010).
%% H.P. Zhang et al., Proc. Natl. Acad. Sci. {\bf 107}, 13626 (2010).

\bibitem{peruani2012} Peruani, F. et al.: Collective motion and nonequilibrium cluster formation in colonies of gliding bacteria. Phys. Rev. Lett. {\bf 108}, 098102 (2012).
%% F. Peruani et al., Phys. Rev. Lett. {\bf 108}, 098102 (2012). 

\bibitem{starruss2012}  Starruss, J. et al.: Pattern-formation mechanisms in motility mutants of Myxococcus xanthus. Interface Focus {\bf 2}, 774-785 (2012).
%% J. Starruss et al., Interface Focus {\bf 2}, 774 (2012). 


\bibitem{schaller10} Schaller, V. et al.: Polar patterns of driven filaments. Nature {\bf 467}, 73-77 (2010).
%% V. Schaller et al., Nature {\bf 467}, 73 (2010).

\bibitem{kudrolli2008} Kudrolli, A. et al.: Swarming and swirling in self-propelled polar granular rods. Phys. Rev. Lett. {\bf 100}, 058001 (2008).
%% A. Kudrolli et al., Phys. Rev. Lett. {\bf 100}, 058001 (2008).

%%\bibitem[{\citenamefont{Kudrolli et~al.}(2006)\citenamefont{Kudrolli, Lumay,
%%  Volfson, and Tsimring}}]{kudrolli2008}
%%\bibinfo{author}{\bibfnamefont{A.}~\bibnamefont{Kudrolli}},
%%  \bibinfo{author}{\bibfnamefont{G.}~\bibnamefont{Lumay}},
%%  \bibinfo{author}{\bibfnamefont{D.}~\bibnamefont{Volfson}}, \bibnamefont{and}
%%  \bibinfo{author}{\bibfnamefont{L.}~\bibnamefont{Tsimring}},
  
 %% Swarming and Swirling in Self-Propelled Polar Granular Rods
% \bibitem{kudrolli08} A. Kudrolli et al., Phys. Rev. Lett. {\bf 100}, 058001
 % (2008); A. Kudrolli, Phys. Rev. Lett. {\bf 104}, 088001 (2010).

\bibitem{deseigne2010} Deseigne, J., Dauchot, O., and Chat\'e, H.: Collective motion of vibrated polar disks. Phys. Rev. Lett. {\bf 105}, 098001 (2010).
%% J. Deseigne, O. Dauchot, and H. Chat\'e, Phys. Rev. Lett. {\bf 105}, 098001 (2010). 

%
%
%\bibitem[{\citenamefont{Deseigne et~al.}(2010)\citenamefont{Deseigne, Dauchot,
%  and Chat{\'e}}}]{deseigne2010}
%\bibinfo{author}{\bibfnamefont{J.}~\bibnamefont{Deseigne}},
%  \bibinfo{author}{\bibfnamefont{O.}~\bibnamefont{Dauchot}}, \bibnamefont{and}
%  \bibinfo{author}{\bibfnamefont{H.}~\bibnamefont{Chat{\'e}}},
%  \bibinfo{journal}{Phys. Rev. Lett.} \textbf{\bibinfo{volume}{105}},
%  \bibinfo{pages}{098001} (\bibinfo{year}{2010}).

\bibitem{weber2013} Weber, C.A. et al.: Long-range ordering of vibrated polar disks. Phys. Rev. Lett. {\bf 110}, 208001 (2013).
%% C. A. Weber, T. Hanke, J. Deseigne, S. Leonard, O. Dauchot, E. Frey, and H. Chat{\'e}, Phys. Rev. Lett. {\bf 110}, 208001 (2013). 

\bibitem{jiang2010} Jiang, H.-R., Yoshinaga, N., and Sano, M.: Active motion of a janus particle by self-thermophoresis in a defocused laser beam. Phys. Rev. Lett. {\bf 105}, 268302 (2010).
%% H.-R. Jiang, N. Yoshinaga, and M. Sano, Phys. Rev. Lett. {\bf 105}, 268302 (2010).

%\bibitem[{\citenamefont{Jiang et~al.}(2010)\citenamefont{Jiang, Yoshinaga, and
%  Sano}}]{jiang2010}
%\bibinfo{author}{\bibfnamefont{H.-R.} \bibnamefont{Jiang}},
%  \bibinfo{author}{\bibfnamefont{N.}~\bibnamefont{Yoshinaga}},
%  \bibnamefont{and} \bibinfo{author}{\bibfnamefont{M.}~\bibnamefont{Sano}},
%  \bibinfo{journal}{Phys. Rev. Lett.} \textbf{\bibinfo{volume}{105}},
%  \bibinfo{pages}{268302} (\bibinfo{year}{2010}).

\bibitem{golestanian2012} Golestanian, R.: Collective behavior of thermally active colloids. Phys. Rev. Lett. {\bf 108}, 038303 (2012).
%% R. Golestanian, Phys. Rev. Lett. {\bf 108}, 038303 (2012).

%\bibitem[{\citenamefont{Golestanian}(2012)}]{golestanian2012}
%\bibinfo{author}{\bibfnamefont{R.}~\bibnamefont{Golestanian}},
%  \bibinfo{journal}{Phys. Rev. Lett.} \textbf{\bibinfo{volume}{108}},
%  \bibinfo{pages}{038303} (\bibinfo{year}{2012}).

%\bibitem[{\citenamefont{Theurkauff et~al.}(2012)\citenamefont{Theurkauff,
%  Cottin-Bizzone, Palacci, Ybert, and Bocquet}}]{theurkauff2012}
%\bibinfo{author}{\bibfnamefont{I.}~\bibnamefont{Theurkauff}},
%  \bibinfo{author}{\bibfnamefont{C.}~\bibnamefont{Cottin-Bizzone}},
%  \bibinfo{author}{\bibfnamefont{J.}~\bibnamefont{Palacci}},
%  \bibinfo{author}{\bibfnamefont{C.}~\bibnamefont{Ybert}}, \bibnamefont{and}
%  \bibinfo{author}{\bibfnamefont{L.}~\bibnamefont{Bocquet}},
%  \bibinfo{journal}{Phys. Rev. Lett.} \textbf{\bibinfo{volume}{108}},
%  \bibinfo{pages}{268303} (\bibinfo{year}{2012}).

\bibitem{theurkauff2012} Theurkauff, C. et al.: Dynamic clustering in active colloidal suspensions with chemical signaling. Phys. Rev. Lett. {\bf 108}, 268303 (2012).
%% Theurkauff et~al., Phys. Rev. Lett. {\bf 108}, 268303 (2012).

\bibitem{palacci2013} Palacci, J. et al.: Living crystals of light-activated colloidal surfers. Science {\bf 339}, 936-940 (2013).
%% Palacci et~al., Science {\bf 339}, 936 (2013).


%\bibitem[{\citenamefont{Palacci et~al.}(2013)\citenamefont{Palacci, Sacanna,
%  Steinberg, Pine, and Chaikin}}]{palacci2013}
%\bibinfo{author}{\bibfnamefont{J.}~\bibnamefont{Palacci}},
%  \bibinfo{author}{\bibfnamefont{S.}~\bibnamefont{Sacanna}},
%  \bibinfo{author}{\bibfnamefont{A.}~\bibnamefont{Steinberg}},
%  \bibinfo{author}{\bibfnamefont{D.}~\bibnamefont{Pine}}, \bibnamefont{and}
%  \bibinfo{author}{\bibfnamefont{P.}~\bibnamefont{Chaikin}},
%  \bibinfo{journal}{Science} \textbf{\bibinfo{volume}{339}},
%  \bibinfo{pages}{936} (\bibinfo{year}{2013}).

\bibitem{paxton2004}  Paxton, W. et al.: Catalytic nanomotors: autonomous movement of striped nanorods. J. Am. Chem. Soc. {\bf 126}, 13424-13431 (2004).
%% W.~Paxton et al., J. Am. Chem. Soc., {\bf 126}, 13424 (2004).

%\bibitem[{\citenamefont{et~al.}(2004)}]{paxton2004}
%\bibinfo{author}{\bibfnamefont{W.~Paxton} \bibnamefont{et~al.}},
%  \bibinfo{journal}{J. Am. Chem. Soc.} \textbf{\bibinfo{volume}{126}},
%  \bibinfo{pages}{13424} (\bibinfo{year}{2004}).

\bibitem{mano2005} Mano, N., and Heller, A.: Bioelectrochemical propulsion. J. Am. Chem. Soc. {\bf 127}, 11574-5 (2005).
%% N. Mano and A. Heller, J. Am. Chem. Soc. {\bf 127}, 11574 (2005). 

%\bibitem[{\citenamefont{Mano and Heller}(2005)}]{mano2005}
%\bibinfo{author}{\bibfnamefont{N.}~\bibnamefont{Mano}} \bibnamefont{and}
%  \bibinfo{author}{\bibfnamefont{A.}~\bibnamefont{Heller}},
%  \bibinfo{journal}{J. Am. Chem. Soc.} \textbf{\bibinfo{volume}{127}},
%  \bibinfo{pages}{11574} (\bibinfo{year}{2005}).

\bibitem{rucker2007}  R\"uckner, G., and Kapral, R.: Chemically powered nanodimers. Phys. Rev. Lett. {\bf 98}, 150603 (2007).
%% G. R{\"u}ckner and R. Kapral, Phys. Rev. Lett. {\bf 98}, 150603 (2007).

\bibitem{howse2007} Howse, J. et al.: Self-motile colloidal particles: from directed propulsion to random walk. Phys. Rev. Lett. {\bf 99}, 048102 (2007).
%%  J. Howse et al., Phys. Rev. Lett. {\bf 99}, 048102 (2007).



%\bibitem[{\citenamefont{R{\"u}ckner and Kapral}(2007)}]{rucker2007}
%\bibinfo{author}{\bibfnamefont{G.}~\bibnamefont{R{\"u}ckner}} \bibnamefont{and}
%  \bibinfo{author}{\bibfnamefont{R.}~\bibnamefont{Kapral}},
%  \bibinfo{journal}{Phys. Rev. Lett.} \textbf{\bibinfo{volume}{98}},
%  \bibinfo{pages}{150603} (\bibinfo{year}{2007}).
%
%\bibitem[{\citenamefont{Howse et~al.}(2007)\citenamefont{Howse, Jones, Ryan,
%  Gough, Vafabakhsh, and Golestanian}}]{howse2007}
%\bibinfo{author}{\bibfnamefont{J.}~\bibnamefont{Howse}},
%  \bibinfo{author}{\bibfnamefont{R.}~\bibnamefont{Jones}},
%  \bibinfo{author}{\bibfnamefont{A.}~\bibnamefont{Ryan}},
%  \bibinfo{author}{\bibfnamefont{T.}~\bibnamefont{Gough}},
%  \bibinfo{author}{\bibfnamefont{R.}~\bibnamefont{Vafabakhsh}},
%  \bibnamefont{and}
%  \bibinfo{author}{\bibfnamefont{R.}~\bibnamefont{Golestanian}},
%  \bibinfo{journal}{Phys. Rev. Lett.} \textbf{\bibinfo{volume}{99}},
%  \bibinfo{pages}{048102} (\bibinfo{year}{2007}).
  
\bibitem{golestanian2007} Golestanian, R., Liverpool, T.B., and Ajdari, A.: Propulsion of a molecular machine by asymmetric distribution of reaction products. Phys. Rev. Lett. {\bf 94}, 220801 (2005).
%% R. Golestanian, T.B. Liverpool, and A. Ajdari, Phys. Rev. Lett. {\bf 94}, 220801 (2005).  
  
%\bibitem[{\citenamefont{R.~Golestanian and Ajdari}(2005)}]{golestanian2007}
%\bibinfo{author}{\bibfnamefont{} \bibnamefont{R.~Golestanian, T.B. Liverpool}}
%  \bibnamefont{and} \bibinfo{author}{\bibfnamefont{A.}~\bibnamefont{Ajdari}},
%  \bibinfo{journal}{Phys. Rev. Lett.} \textbf{\bibinfo{volume}{94}},
%  \bibinfo{pages}{220801} (\bibinfo{year}{2005}).

\bibitem{bartolo2013}  Bricard, A. et al.: Emergence of macroscopic directed motion in populations of motile colloids. Nature {\bf 503}, 95-98 (2013).
%% A. Bricard et al., Nature {\bf503}, 95 (2013).
%Antoine Bricard, Jean-Baptiste Caussin, Nicolas Desreumaux, Olivier Dauchot, Denis Bartolo, Nature \bf{503},
 %   95 (2013).

\bibitem{thutupalli2011} Thutupalli, S., Seemann, R., and Herminghaus S.: Swarming behavior of simple model squirmers. New J. Phys. {\bf 13}, 073021 (2011).
%% S. Thutupalli, R. Seemann, S. Herminghaus, New J. Phys. {\bf13}, 073021 (2011). 

%%%%% Granular media

%%%%%%%%%%%%%%%%%%%%%%%%%%%%%
\bibitem{vicsek1995} Vicsek, T. et al.: Novel type of phase transition in a system of self-driven particles. Phys. Rev. Lett. {\bf 75}, 1226 (1995).
%% T. Vicsek et al., Phys. Rev. Lett. {\bf 75}, 1226 (1995).

\bibitem{gregoire2004} Gr\'egoire, G., and Chat\'e, H.: Onset of collective and cohesive motion. Phys. Rev. Lett. {\bf 92}, 025702 (2004).
%% G. Gr\'egoire and H. Chat\'e, Phys. Rev. Lett. {\bf 92}, 025702 (2004).
    
\bibitem{peruani2008} Peruani, F., Deutsch, A., and B\"ar, M.: A mean-field theory for self-propelled particles interacting by velocity alignment mechanisms. Eur. Phys. J. Special Topics {\bf 157}, 111-122 (2008); Ginelli, F. et al.: Large-scale collective properties of self-propelled rods. Phys. Rev. Lett. {\bf 104}, 184502 (2010).
%% F. Peruani, A. Deutsch, and M. B\"ar, Eur. Phys. J-Spec. Top.,{\bf 157}, 111 (2008); 
%% F. Ginelli et al., Phys. Rev. Lett. {\bf 104}, 184502  (2010). 

\bibitem{peruani2006} Peruani, F., Deutsch, A., and B\"ar, M.: Nonequilibirum clustering of self-propelled rods. Phys. Rev. E {\bf 74}, 030904(R) (2006).

\bibitem{chate2006} Chat\'e, H., Ginelli, F., and Montagne, R.: Simple model for active nematics: quasi-long-range order and giant fluctuations. Phys. Rev. Lett. {\bf 96}, 180602 (2006).
%% H. Chat\'e, F. Ginelli and R. Montagne, Phys. Rev. Lett. {\bf 96}, 180602 (2006).

\bibitem{bussemaker1997}  Bussemaker, H.J., Deutsch, A., and Geigant, E.: Mean-field analysis of a dynamical phase transition in a cellular automaton model for collective motion. Phys. Rev. Lett. {\bf 78}, 5018 (1997).
%% H.J. Bussemaker, A. Deutsch, and E. Geigant, Phys. Rev. Lett. {\bf 78}, 5018 (1997).

\bibitem{csahok1995} Csah\'ok, Z., and Vicsek, T.: Lattice-gas model for collective biological motion. Phys. Rev. E {\bf 52}, 5297--5303 (1995).
%% Z. Csah\'ok and T. Vicsek, Phys. Rev. E {\bf 52}, 5297 (1995). 

%%% Lattice swarming models in 1D: transition second order, though not clear related to orientation

\bibitem{loan1999} O'Loan, O.J., and Evans, M.R.: Alternating steady state in one-dimensional flocking. J. Phys. A: Math. Gen. {\bf 32}, 99 (1999).
%% O.J. O'Loan and M.R. Evans, J. Phys. A: Math. Gen. {\bf 32}, 99 (1999).

\bibitem{raymond2006} Raymond, J.R., and Evans, M.R.: Flocking regimes in a simple lattice model. Phys. Rev. E {\bf 73}, 036112 (1--13) (2006).
%% J.R. Raymond and M.R. Evans, Phys. Rev. E {\bf 73}, 036112 (2006).

\bibitem{chepizkho2013a}  Chepizhko, O., Altmann, E., and Peruani, F.: Optimal noise maximizes collective motion in heterogeneous media. Phys. Rev. Lett. {\bf 110}, 238101 (2013).
%% O. Chepizhko. E. Altmann, and F. Peruani Phys. Rev. Lett. 110, 238101 (2013). 

\bibitem{chepizkho2013b} Chepizhko, O., and Peruani, F.: Diffusion, subdiffusion, and trapping of active particles in heterogeneous media. Phys. Rev. Lett. {\bf 111}, 160604 (2013).
%% O. Chepizhko and F. Peruani Phys. Rev. Lett. 111, 160604 (2013). 

\bibitem{reichhardt2014} Reichhardt, C., and Olson Reichhardt, C.J.: Active matter transport and jamming on disordered landscapes. arXiv:1402.3260 (2014).
%% C. Reichhardt and C.J. Olson Reichhardt, arXiv:1402.3260 (2014).

\bibitem{quint2014}  	Quint, D.A., and Gopinathan, A.: Swarming in disordered environments. arXiv:1302.6564 (2013).
%% David A. Quint, Ajay Gopinathan, arXiv:1302.6564 (2013). 

\bibitem{peruani2011} Peruani, F. et al.: Traffic jams, gliders, and bands in the quest of collective motion of self-propelled particles. Phys. Rev. Lett. {\bf 106}, 128101 (2011).
%% F. Peruani et al., Phys. Rev. Lett. {\bf 106}, 128101 (2011). 

\bibitem{thompson2011} Thompson A.G. et al.: Lattice models of nonequilibrium bacterial dynamics. J. Stat. Mech. {\bf 11}, P02029 (2011).
%% A.G. Thompson et al., J. Stat. Mech. {\bf 11}, P02029 (2011). 

\bibitem{tailleur2012} Farrell, F. D. C. et al.: Pattern formation in self-propelled particles with density-dependente motiliy. Phys. Rev. Lett. {\bf 108}, 248101 (2012).
%%  F. D. C. Farrell et al., Phys. Rev. Lett. {\bf 108}, 248101(2012). 

\bibitem{tailleur2008} Tailleur, J., and Cates, M. E.: Statistical mechanics of interacting run-and-tumble bacteria. Phys. Rev. Lett. {\bf 100}, 218103 (2008).
%% J. Tailleur and M. E. Cates, Phys. Rev. Lett. {\bf 100}, 218103 (2008).

\bibitem{tailleur2013} Cates, M.E., and Tailleur, J.: When are active Brownian particles and run-and-tumble particles equivalent? 
Consequences for motility-iniduced phase separation. Europhys. Lett. {\bf 101}, 20010 (2013).
%% M.E. Cates and J. Tailleur, Europhys. Lett. {\bf 101}, 20010 (2013). 

\bibitem{fily2012} Fily, Y., and Marchetti, M.C.: Athermal phase separation of self-propelled particles with no alignment. Phys. Rev. Lett. {\bf 108}, 235702 (2012).
%% Y. Fily and M.C. Marchetti, Phys. Rev. Lett. {\bf 108}, 235702 (2012). 

\bibitem{redner2013} Redner, G., Hagan, M.F., and Baskaran, A.: Structure and dynamics of a phase-separating active colloidal fluid. Phys. Rev. Lett. {\bf 110}, 055701 (2013).
%% G. Redner, M.F. Hagan, A. Baskaran, Phys. Rev. Lett. {\bf 110}, 055701 (2013).

\bibitem{fily2014} Fily, Y., Henkes, S., and Marchetti, M.C.: Freezing and phase separation of self-propelled disks. Soft Matter {\bf 10}, 2132-2140 (2014).
%% Y. Fily, S. Henkes, and M.C. Marchetti, Soft Matter (2014).

\bibitem{valeriani2013}  Mognetti, B.M. et al.: Living clusters and crystals from low-density suspensions of active colloids. Phys. Rev. Lett. {\bf 111}, 245702 (2013).
%% B.M. Mognetti et al. Phys. Rev. Lett. {\bf 111}, 245702 (2013).

\bibitem{wysocki2013}  Wysocki, A., Winkler, R.G., and Gompper, G.: Cooperative motion of active brownian spheres in three-dimensional dense suspensions. arXiv:1308.6423 (2013)
%% A. Wysocki, R.G. Winkler, G. Gompper, arXiv:1308.6423 (2013)

\bibitem{cates2013} Stenhammar, J. et al.: Continuum theory of phase separation kinectis for active brownian particles. Phys. Rev. Lett. {\bf 111}, 145702 (2013).
%% J. Stenhammar et al., Phys. Rev. Lett. {\bf 111}, 145702 (2013).

\bibitem{speck2013} Speck, T. et al.: Effective Cahn-Hilliard equation for phase separation of active Brownian particles. arXiv:1312.7242 (2013); Bialk\'e, J., Lšwen, H., and Speck, T.: Microscopic theory for the phase separation of self-propelled repulsive disks. Europhys. Lett. 103, 30008 (2013).
%% T. Speck et al., arXiv:1312.7242 (2013); J. Bialk{\'e}, H. L{\"o}wen, and T. Speck, Europhys. Lett. {\bf 103} 30008 (2013).

\bibitem{toner1995} Toner, J., and Tu, Y.: Long-range order in a two-dimensional dynamical XY model: how birds fly together. Phys. Rev. Lett. {\bf 75}, 4326 (1995).
%% J. Toner and Y. Tu, Phys. Rev. Lett. {\bf 75}, 4326 (1995). 

\bibitem{toner1998} Toner, J., and Tu, Y.: Flocks, herds, and schools: a quantitative theory of flocking. Phys. Rev. E {\bf 58}, 4828--4858 (1998).
%% J. Toner and Y. Tu, Phys. Rev. E {\bf 58}, 4828 (1998).

\bibitem{bertin2009} Bertin, E., Droz, M., and Gr\'egoire, G.: Hydrodynamic equations for self-propelled particles: microscopic derivation and stability analysis. J. Phys. A: Math. Theor. {\bf 42}, 445001 (2009).
%% E. Bertin, M. Droz and G. Gr\'egoire J. Phys. A 42, 445001 (2009).

\bibitem{mishra2010} Mishra, S., Baskaran, A., and Marchetti, M.C.: Fluctuations and pattern formation in self-propelled particles. Phys. Rev. E {\bf 81}, 061916 (1--14) (2010).
%% S. Mishra, A. Baskaran, and M.C. Marchetti, Phys. Rev. E 81, 061916 (2010). 

\bibitem{caussin2014} Caussin, J-B. et al.: Emergent spatial structures in flocking models: a dynamical system insight. arXiv:1401.1315 (2014). 
%% J-B. Caussin et al.,  arXiv:1401.1315 (2014).

\bibitem{marchetti2013} Marchetti, M. C. et al.: Hydrodynamics of soft active matter. Rev. Mod. Phys. {\bf 85}, 1143--1189 (2013). 
%% M. C. Marchetti et al., Rev. Mod. Phys. {\bf 85}, 1143 (2013) 

\bibitem{peshkov2012} Peshkov, A. et al.: Nonlinear field equations for aligning self-propelled rods. Phys. Rev. Lett. {\bf 109}, 268701 (2012).
%% A. Peshkov et al., Phys. Rev. Lett. {\bf 109}, 268701 (2012). 

\bibitem{ihle2011}  Ihle, T.: Kinetic theory of flocking: derivation of hydrodynamic equations. Phys. Rev. E {\bf 83}, 030901(R) (2011).
%% T. Ihle, Phys. Rev. E {\bf 83}, 030901 (2011). 

\bibitem{degond2008} Degond, P., and Motsch, S.: Continuum limit of self-driven particles with orientation interaction. Math. Models Methods Appl. Sci. {\bf 18}, 1193-1215 (2008).
%% P. Degond and S. Motsch, Math. Models Methods Appl. Sci. {\bf 18}, 1193 (2008). 

\bibitem{soto2014} Soto, R., and Golestanian, R.: Run-and-tumble dynamics in a crowded environment: persistent exclusion process for swimmers. Phys. Rev. E {\bf 89}, 012706 (1--7) (2014).
%% R. Soto and R. Golestanian, Phys. Rev. E {\bf 89}, 012706 (2014).

\bibitem{Dean96} Dean, D.S.: Langevin equation for the density of a system of interacting Langevin processes. J. Phys. A: Math. Gen. {\bf 29}, L613-L617  (1996). 
%% D.S.  Dean, J. Phys. A: Math. Gen. {\bf 29}, L613 (1996).

%% \bibitem{Dawson} Dawson-Gartner......

\bibitem{gardiner} Gardiner, C. W.: Handbook of stochastic methods (Springer, Heildelberg, 2004).
%% C. W. Gardiner, {\it Handbook of stochastic methods}  (Springer, Heildelberg, 2004).

\bibitem{peruani2013} Peruani, F., and B\"ar, M.: A kinetic model and scaling properties of non-equilibrium clustering of self-propelled particles. New J. Phys. 15, 065009 (2013). 
%% F. Peruani and M. B{\"a}r, New J. Phys. {\bf 15}, 065009 (2013).

% \bibitem{MFT02} L. Bertini et al. J Stat Mech 2002
%% \bibitem{DawsonGartner} D.A. Dawson and J. G\"artner. Stochastics: formerly Stochastics and Stochastics Reports, 20:4, 247-308 (1987)

\end{thebibliography}
\end{document}